\documentclass[%
 aip,
 amsmath,amssymb,
 reprint,%
]{revtex4-1}

\usepackage{graphicx}
\usepackage{dcolumn}
\usepackage{bm}
\usepackage{float}
\usepackage{xcolor}

\usepackage[utf8]{inputenc}
\usepackage[T1]{fontenc}
\usepackage{mathptmx}
\usepackage{etoolbox}

\makeatletter
\def\@email#1#2{%
 \endgroup
 \patchcmd{\titleblock@produce}
  {\frontmatter@RRAPformat}
  {\frontmatter@RRAPformat{\produce@RRAP{*#1\href{mailto:#2}{#2}}}\frontmatter@RRAPformat}
  {}{}
}%
\makeatother
\begin{document}

\preprint{AIP/123-QED}

\title[Sample title]{Mathematical model of fluid front dynamics driven by porous media pumps}

\author{Andreu Benavent-Claró$^*$}%
 \affiliation{Condensed Matter Physics Department, Physics Faculty, University of Barcelona, Barcelona, Spain.}
 \affiliation{Institute of Nanoscience and Nanotecnology (IN2UB), University of Barcelona, Barcelona, Spain.}
 \email{abenavent@ub.edu}
\author{Yara Alvarez-Braña}%
\affiliation{Microfluidics Cluster UPV/EHU, BIOMICs Microfluidics Group, University of the Basque Country UPV/EHU, Vitoria-Gasteiz, Spain.}
\affiliation{Microfluidics Cluster UPV/EHU, Analytical Microsystems \& Materials for Lab-on-a-Chip (AMMa-LOAC) Group, University of the Basque Country UPV/EHU, Vitoria-Gasteiz, Spain.}
\author{Fernando Benito-Lopez}
\affiliation{Microfluidics Cluster UPV/EHU, Analytical Microsystems \& Materials for Lab-on-a-Chip (AMMa-LOAC) Group, University of the Basque Country UPV/EHU, Vitoria-Gasteiz, Spain.}
\author{Lourdes Basabe-Desmonts}
\affiliation{Microfluidics Cluster UPV/EHU, BIOMICs Microfluidics Group, University of the Basque Country UPV/EHU, Vitoria-Gasteiz, Spain.}
\affiliation{Basque Foundation of Science, IKERBASQUE, Vitoria-Gasteiz, Spain.}
\author{Aurora Hernandez-Machado}
\affiliation{Condensed Matter Physics Department, Physics Faculty, University of Barcelona, Barcelona Spain.}
\affiliation{Institute of Nanoscience and Nanotecnology (IN2UB), University of Barcelona, Barcelona, Spain.}

\date{\today}

\begin{abstract}

Air-permeable porous media hosts air within their pores. Upon removal from the interior of the material, these porous media have the tendency to reabsorb air from the surrounding, acting as a suction pump. Therefore, the technique used to convert porous media into a pump, consists of degassing the material to remove their air inside. The suction property when recovering the air, can be used to move a liquid through a microfluidic channel. Porous media pumps are very accurate devices to move liquids in a completely controlled way. {By studying the dynamics of the liquid front moved by these pumps, it is possible to extract characteristic properties of both the fluid and the porous material.} In this article, we have developed a theoretical mathematical model that precisely characterizes the dynamics of a liquid moved by a degassed porous media pump through a microchannel by comparing it with experimental data. {We have seen the differences between sealing the external surface of the pump so that it cannot absorb air from the outside, both mathematically and experimentally.} We have observed that, in all cases, the theory fits satisfactorily with the experiments, corroborating the validity of the model. The creation of microfluidic pumps using porous media can be a very useful tool in various fields due to its long operating time, small size and the fact that it operates without any external power source.
\end{abstract}

\maketitle

\section{INTRODUCTION}

The study and development of self-powered micropumps at the microscale is an area of interest nowadays due to its important applications, such as autonomous microsystems and the power that they showcase despite their small size, making this type of pump completely portable. When placed in a microfluidic system, these micropumps produce a pressure difference inside of the channel capable of moving a fluid without any bulky external power source \cite{nowoo}.

Many self-powered autonomous pumps have been developed through various approaches and for different purposes \cite{3,4,5}, which can be required for many microfluidic applications. For instance, the development of a macroscale chemostat for {\it E. coli} population control studying an artificial microbial ecosystem \cite{6,7}, cancer monitoring \cite{8,9}, real-time changes in gene expression in budding yeast \cite{10}, the measurement of the response over time of cellular lines to the administration of drug-containing media \cite{11,12}, among other important applications \cite{13,14,15,16,17,18,19,20,21,22}.

\begin{figure}
    \centering
    \includegraphics[width=0.5\textwidth]{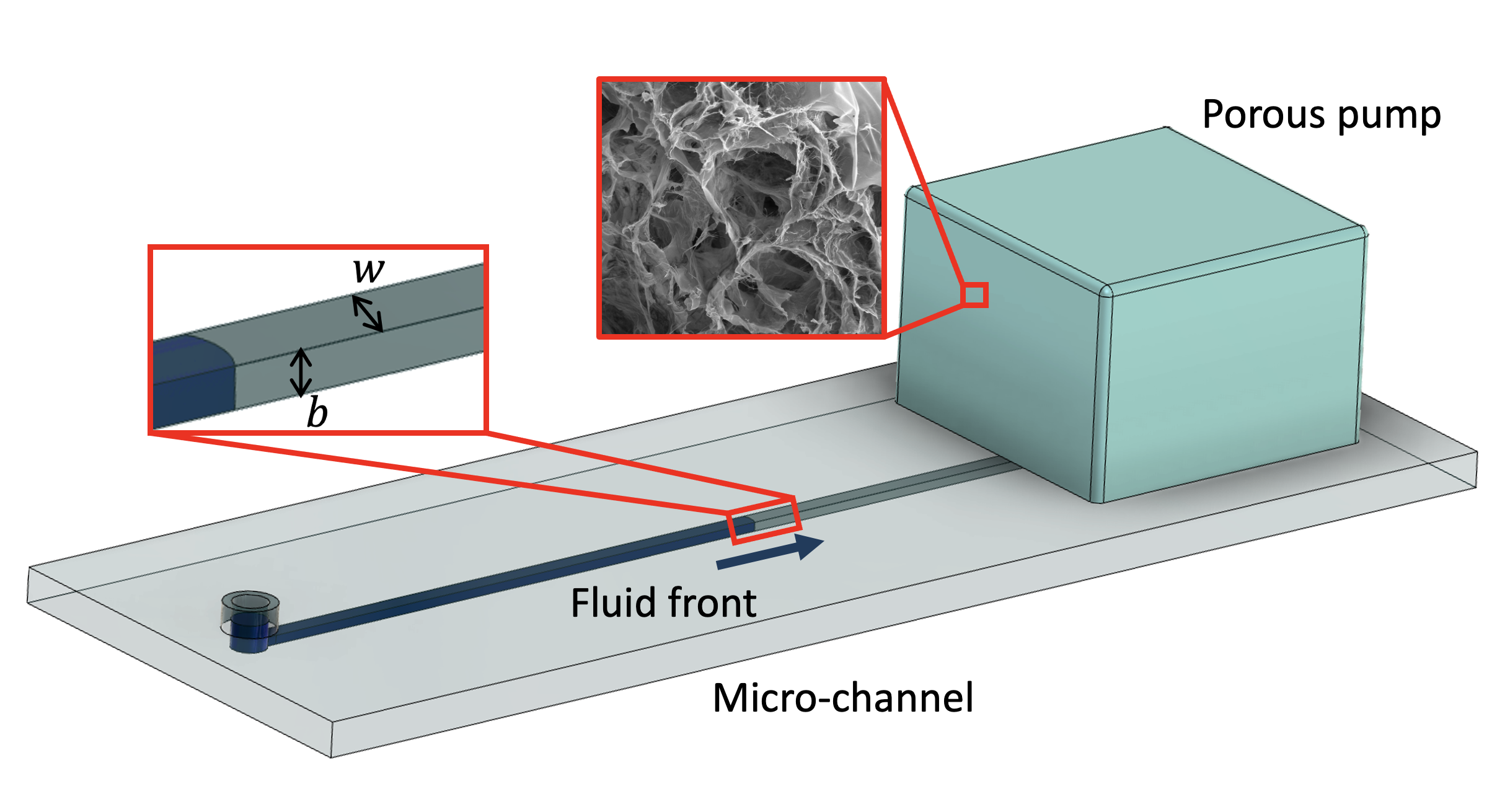}
    \caption{Schematic representation of a porous pump suctioning air to move a liquid column through a channel.}
    \label{fig.channel}
\end{figure}

The most promising type of pump that meets the requirements of being small, self-contained, portable and capable of pulling liquids for long periods of time are porous media pumps. This pumps are composed of an air-permeable porous media\cite{woo,jaione,yaravell}. These materials possess air within their internal structure. The amount of air retained inside the pump depends on the environmental pressure in which it is located and the pump volume. When degassing the material in a vacuum chamber, it loses a fraction of the air contained due to the pressure gradient. Once the pump is placed at atmospheric pressure again, it recovers the air it lost from the environment. Depending on the properties of the pump pore, {the surface of the pump and the air removed from it in the degassing process,} it will take more or less time to recover this air. This air suction effect can be used to turn a porous medium into a suction pump. Thus, by placing the material in the outlet of a fluidic microchannel, the recovered air creates a pressure gradient capable of moving a fluid through the channel\cite{jaione,woo}, as can be seen in Figure \ref{fig.channel}.

Any air-permeable porous medium can be used to create a porous medium pump. However, to prevent air reabsorption from occurring almost instantaneously, the pores of the medium must be small enough to allow prolonged air diffusion over time. The most common material used for this purpose is polydimethylsiloxane (PDMS)\cite{jaione,woo,nowoo,zhao}. Nevertheless, it is not the only one, as there are countless other materials that have the same characteristics, such as some 3D printing resins\cite{yaravell}.

{While significant advances have been made in the development of porous media pumps, existing studies have focused on specific materials, such as PDMS, developing empirical models focused on the particular system under study. Although these studies have successfully demonstrated the utility of these types of pumps, they do not provide a general model to predict the behavior of porous media pumps regardless of the pump material and the sample fluid being moved.}

{In addition, experimental evidence suggests that coating the outer surface of the pump with an air-impermeable material significantly improves its efficiency by channeling air absorption exclusively through the effective surface \cite{zhao}. However, a detailed theoretical understanding of this effect has not yet been established. By incorporating porous effects into the analysis, it becomes possible to build a robust mathematical model capable of describing the behavior of porous media pumps in general. This knowledge not only demonstrates the versatility of using various porous media to create efficient pumps, but also provides a solid basis for optimizing their design and expanding their range of applications.}

\begin{figure}
    \centering
    \includegraphics[width=0.5\textwidth]{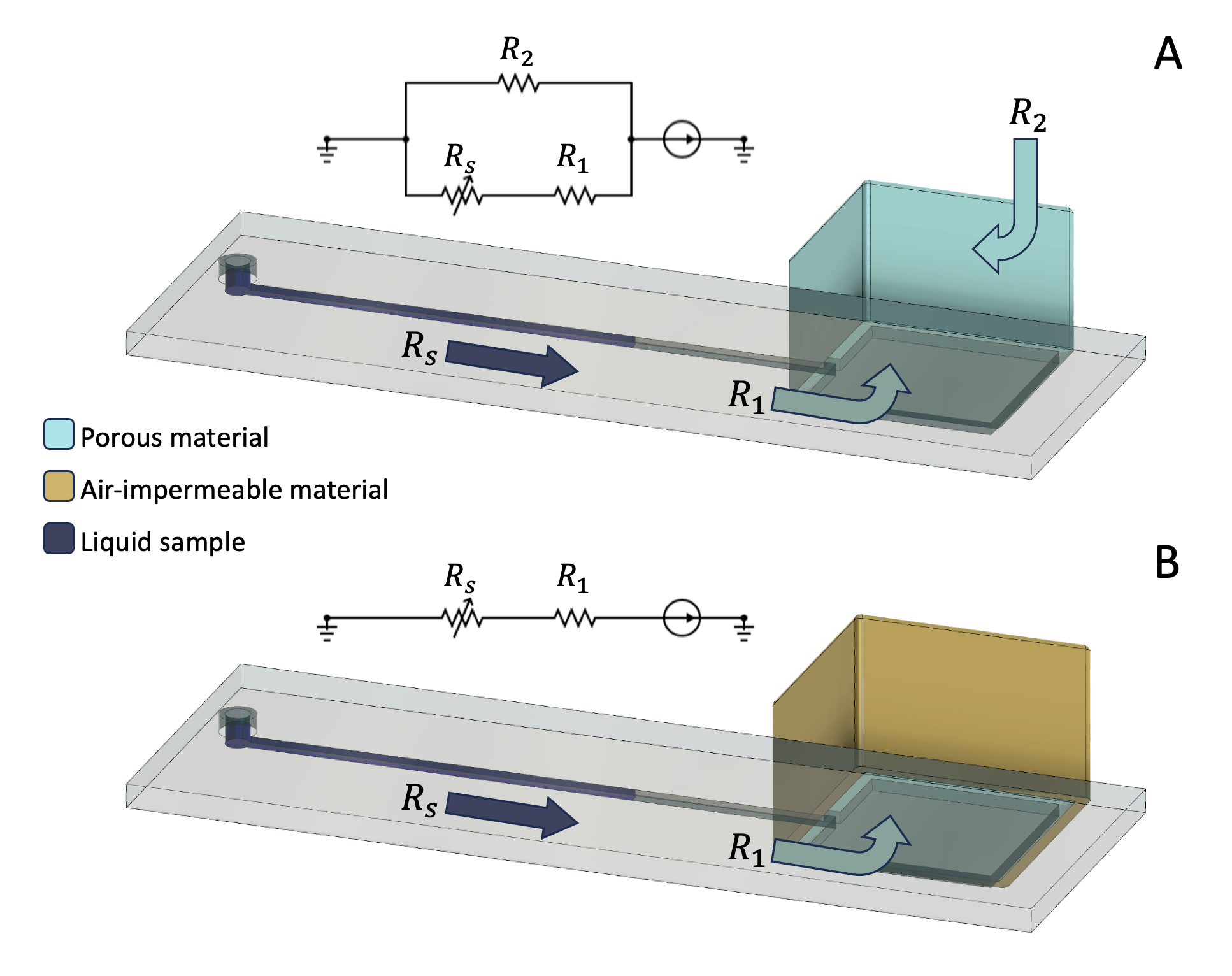}
    \caption{Difference between a porous pump with an air-impermeable coating and a uncoated pump. The arrows shown the fluidic resistances present in each system, a schematic of the resistances considered is also shown.}
    \label{fig.coated}
\end{figure}

{In this work, we derive a general mathematical model that predicts the behavior of any liquid moved by any porous media pump, independent of the material of the pump used. Additionally, we analyze the role of coating porous pumps in enhancing efficiency by focusing air reabsorption exclusively through the microchannel. Figure \ref{fig.coated}A illustrates the behavior of an uncoated pump, which absorbs air from all surfaces, while Figure \ref{fig.coated}B shows a coated configuration where air absorption is limited to the channel, enhancing the suction effect. This work bridges the existing theoretical gaps and demonstrates the broader applicability of porous media pumps in microfluidic systems.}

\section{MATHEMATICAL MODEL}
In this theoretical study, we describe the behavior of a porous media pump using the principles of parallel and series fluidic resistances. This model can describe both the behavior of those pumps that present all external surfaces covered with an impermeable material and the uncovered ones. Figure \ref{fig.coated} shows the air flow differences between {an uncoated (Figure \ref{fig.coated}A) and a coated (Figure \ref{fig.coated}B) pump.}

Considering that the absorption of air by the porous media is not instantaneous, it can be inferred that the porous material generates resistance to the flow of air through its cavities. Thus, in the microfluidic system composed of a suction pump and a channel we will have different fluidic resistances: one caused by the fluid flow when moving through the microchannel and the others due to the absorption of the air in the porous media.

In the general case we will have three different fluidic resistances. The first one is due to the viscous liquid moving in the channel ($R_s$). This resistance increases as a function of the flow rate of the liquid in the channel. Therefore, the more volume of liquid displaced, the greater this resistance will be. The expression of this resistance is given by:
\begin{eqnarray}
    R_s(t)=\frac{12\eta_s}{b^2S_s^2}V(t)\equiv\xi_sV(t),
    \label{eq.rsamp}
\end{eqnarray}
where $\eta_s$ is the viscosity of the liquid sample; $V(t)$, volume of liquid displaced by the pump; $b$, the height of the channel; $S_s$ the channel cross section; and $\xi_s$ is a constant defined as $\xi_s=\frac{12\eta_s}{b^2S_s^2}$.

On the other hand, there are resistances caused by air flowing through the porous media. In this case, two resistances can be identified. The first, {$R_1$}, is due to air from the channel entering the pump through its effective surface, which corresponds to the portion of the pump in contact with the channel (Figure 2B).The second, $R_2$, arises from air absorption from the environment through the external surfaces of the pump (Figure 2A). Both resistances follow the same mathematical expression:
\begin{eqnarray}
    R_i = \frac{\eta_aL}{K_iS_i},
    \label{eq.rpdms}
\end{eqnarray}
where $\eta_a$ is the viscosity of air; $L$, a characteristic length of the porous media; $S_i$, the surface of the porous media pump ($S_1$, corresponds to the effective surface and $S_2$ to the external surface of the pump); and $K_i$, the permeability of the porous media. {Permeability is an unchanging characteristic property of an specific porous media. It depends on the porous geometry and the number of porous. It is related to the resistance that a fluid generates when flowing through it by equation \eqref{eq.rpdms}. If we coat the pump with a material of different permeability, the permeability of the system will change. For instance, if we coat the pump } with an air-impermeable material $K=0$.

We consider that, at very low velocities, the air is not compressed and therefore, it does not change its volume. With this consideration, we can use the principle of conservation of mass, obtaining that the flow rate that passes through the fluidic resistor ($R_s$) must be equal to the flow rate that passes through the effective surface resistor ($R_1$).

As far as pressure is concerned, we assumed that the time-dependent pressure gradient of the porous media pump obeys a decay law equation\cite{woo}:
\begin{eqnarray}
    \frac{dP(t)}{dt}=-\frac{1}{\tau}P(t),
\end{eqnarray}
{which solution is:
\begin{eqnarray}
     P(t) = P_0e^{-\frac{t}{\tau}},
     \label{eq.difP}
\end{eqnarray}}
where $P$ is the gauge pressure (the absolute pressure minus the atmospheric pressure); $P_0$, the initial pressure; and $\tau$ is a characteristic time constant.

The system consists of two resistors in series and one in parallel. The liquid flowing through the microchannel generates a resistance ($R_s$) that is in series with the resistance due to the air of the channel flowing through the porous media ($R_1$), both having the same air flow rate, while the resistance due to the air flowing over the external surface ($R_2$) is located in parallel to those previous resistances.

Thus, there will be the same pressure drop across resistor $R_2$ as across the set of resistors $R_s$ and $R_1$, while the flow absorbed by the pump through the channel will be different from the flow absorbed by the external surface. As we can only measure the flow of the channel by observing the fluid front motion, we will define it as $Q(t)$. This value corresponds to the air flow rate absorbed by the channel. 

With these considerations, and using Ohm's law for fluids, we can find a relation between the pressure and the flow rate of the microchannel:
\begin{eqnarray}
    P(t) = (R_s(t)+R_1)Q(t) = \left(\xi_sV(t)+R_1\right)\Dot{V}(t),
    \label{eq.coatdifeq}
\end{eqnarray}
where the flow rate is defined as the temporal derivative of the volume absorbed by the pump. Then, $V(t)$ is the volume of air from the channel absorbed by the pump, or the volume of liquid that advances through the channel.

Solving this differential equation by considering the expression (\ref{eq.difP}) for the pressure, and taking into account that no air volume is absorbed at time zero [$V(t=0)=0$], we obtain the air volume absorbed by the pump from the channel as a function of time:
\begin{eqnarray}
    \frac{\xi_s}{2}V^2(t)+R_1V(t) = P_0\tau\left(1-e^{-\frac{t}{\tau}}\right).
    \label{eq.xcot}
\end{eqnarray}
Equation (\ref{eq.xcot}) is a quadratic polynomial equation for the volume of liquid displaced as a function of an exponential growth for time, whose solution is:

\begin{eqnarray}
    V(t) = \frac{R_1}{\xi_s}\left(\sqrt{1+\frac{2P_0\tau\xi_s}{R_1^2}\left(1-e^{-\frac{t}{\tau}}\right)}-1\right).
\end{eqnarray}

In all cases, the fluid volume displaced stops at large times. The total volume displaced, $V_0=V(t\rightarrow\infty)$, is:
\begin{eqnarray}
    V_0 = \frac{R_1}{\xi_s}\left(\sqrt{1+\frac{2P_0\tau\xi_s}{R_1^2}}-1\right).
    \label{eq.finalV}
\end{eqnarray}
Equation (\ref{eq.xcot}) is linear at low times and quadratic at large times. Then, it presents a crossover between these two regimes. This crossover can be characterized defining a crossover time ($t_{co}$), which tells us when the linear term no longer dominates:
\begin{eqnarray}
    t_{co}=\tau\ln\left(\frac{P_0\tau\xi_s}{P_0\tau\xi_s-4R_1^2}\right).
    \label{eq.tco}
\end{eqnarray}
If the crossover time is greater then the characteristic time constant $\tau$, the linear term dominates. However, if the crossover time is smaller than the characteristic time, it is the quadratic term that dominates.

The flow rate of the fluid moved by the pump is then the derivative of the displaced volume defined as:
\begin{eqnarray}
    Q(t) = \Dot{V}(t) = \frac{P_0}{R_1}\frac{e^{-\frac{t}{\tau}}}{\sqrt{1+\frac{2P_0\tau\xi_s}{R_1^2}\left(1-e^{-\frac{t}{\tau}}\right)}}.
    \label{eq.flowcoated}
\end{eqnarray}

On the other hand, we will define $Q_2$ as the air flow rate that the pump absorbs from the environment. As in the previous case, using the Ohm's law for fluids we can find the volume absorbed by the pump from the environment ($V_2$):
\begin{eqnarray}
    P(t) = R_2Q_2(t) = R_2\Dot{V}_2(t),
\end{eqnarray}
Then, by solving this differential equation we obtain the following expression:
\begin{eqnarray}
    V_2(t) = \frac{P_0\tau}{R_2}\left(1-e^{-\frac{t}{\tau}}\right).
    \label{eq.v2}
\end{eqnarray}

Knowing that the total volume of air absorbed by the pump as a function of time is the sum of the volume absorbed from the channel plus the volume absorbed from the outside, $V_T(t)=V(t)+V_2(t)$, we can find an expression for the total air volume absorbed as a function of time:
\begin{eqnarray}
    V_T(t)=\frac{P_0\tau}{R_2}\left(1-e^{-\frac{t}{\tau}}\right)+\frac{R_1}{\xi_s}\left(\sqrt{1+\frac{2P_0\tau\xi_s}{R_1^2}\left(1-e^{-\frac{t}{\tau}}\right)}-1\right).
\end{eqnarray}
The total volume absorbed by the pump at late times will correspond to the volume of air extracted from the pump in the degassing process. Thus, the volume of air removed from the pump during degassing ($V_{des}= V_T(t\rightarrow\infty)$) is given by:
\begin{eqnarray}
    V_{des} =\frac{P_0\tau }{R_2}+\frac{R_1}{\xi_s}\left(\sqrt{1+\frac{2P_0\tau\xi_s}{R_1^2}}-1\right).
    \label{eq.vdes}
\end{eqnarray}
From this expression, we can find the value of the characteristic time constant $\tau$:
\begin{eqnarray}
    2\left(\frac{R_1+R_2}{\xi_sR_2}+\frac{V_{des}}{R_2}\right)\tau-\frac{P_0}{R_2^2}\tau^2=\frac{V_{des}}{P_0}\left(\frac{2R_1}{\xi_s}+V_{des}\right).
    \label{eq.tau}
\end{eqnarray}
The characteristic time constant $\tau$ is an important parameter since it does not only give us information on the time the pump is actuating, but it also dictates the shape at which the volume of fluid sample is advancing as a function of time as found in Equation (\ref{eq.tco}).

Coated porous media pumps are those pumps that present an air-impermeable coating in its external surface while presenting a permeable behavior on the surface that it is in contact with the channel. Mathematically, this means that the external surface permeability is zero ($K_2=0$), which translates to (see Equation (\ref{eq.rpdms})) an infinitely resistance of the external walls of the porous media pump ($R_2\rightarrow\infty$). This means that no air enters through the outer layers, resulting in $V_2=0$. The expression for the time dependent constant is then:
\begin{eqnarray}
    \tau_{coated} = \frac{V_{des}}{P_0}\left(R_1+\frac{\xi_s}{2}V_{des}\right),
    \label{eq.tauimp}
\end{eqnarray}
where $\tau_{coated}$ is the time constant attributed to a pump coated with an impermeable material.

It can be proved that $\tau_{coated}$ will always be greater than $\tau$ for any $R_2$. This result affirms that the activity of the coated pumps will be much more prolonged in time than the uncoated pumps.

{This mathematical model is valid for any Newtonian sample fluid of viscosity $\eta_s$ and any porous media pump of permeability $K_i$.}

\section{Experimental results and discussion}

To validate the mathematical model presented in this article, several experiments were performed to see the agreement between the model and the experimental data. These experiments consisted in allowing a viscous fluid (water) to flow through a rectangular cross-section channel pulled by a porous media pump: a coated and a uncoated one. Both of them were degassed during different times.

\subsection{Methodology}
As an air porous material to perform the pumps, we have used Polydimethylsiloxane (PDMS). PDMS is a widely used silicone-based polymer known for its porosity. PDMS forms a polymerical structure with interconnected {microscopic} pores that can hold air or other gases in its internal structure. {PDMS has an optimal permeability due to the size and number of pores in its structure for use as a porous media pump because, although the model accepts any type of porous media, if the resistance of the air flowing through it is negligible with respect to the resistance of the liquid sample in the channel, no displacement of the sample would be appreciated.} 

\begin{figure} [h]
    \centering
    \includegraphics[width=0.45\textwidth]{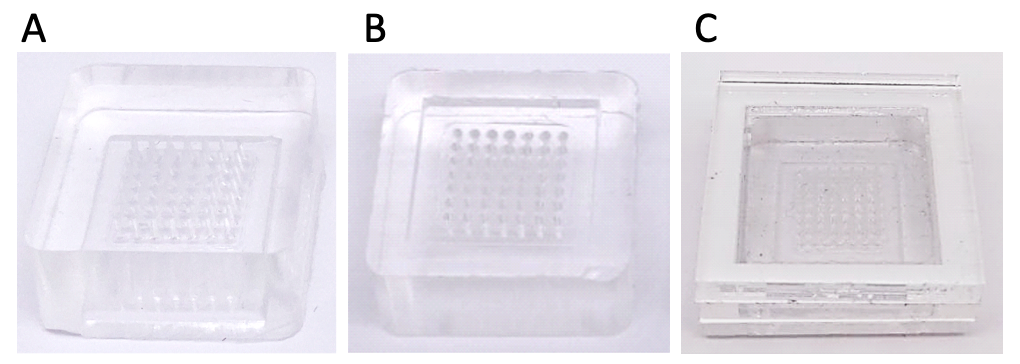}
    \caption{Photograph of the PDMS pumps used. A: top view. B: bottom view. C: epoxy-coated pump. {In all cases it can be seen the cavities in the effective surface designed to increase the effective surface area of the pump and make them more efficient.}}
    \label{fig.bomba}
\end{figure}

Figure \ref{fig.bomba} shows a photograph of the pumps used.
The PDMS pumps used in this article have a cubic shape, {measuring 22×22×9 mm (L×W×H)}, with cavities on the surface in contact with the channel. {These cavities, which can be seen in Figure \ref{fig.bomba}, are a pump design that only serve to increase the effective surface area of the pump (the surface area of the pump in contact with the channel where the sample liquid flows). This increase of the effective surface area serves to observe a higher pump efficiency in the uncoated case.} The pumps have been made from a 3D printing mold, where the liquid PDMS has been poured and polymerized in an oven at 80ºC for two hours.

To create air-impermeable coated porous media pumps for the experiments, we used epoxy resin (Epoxy Araldite, Ceys), a polymer that does not exhibit porosity and, therefore, cannot retain air like PDMS. This makes it an effective air-impermeable material. The entire surface of the porous media pump that was not in contact with the channel was coated with this polymer.

Since the pumps can move a large amount of volume, especially the coated ones, the microchannel where the liquid passes through was designed in the form of a serpentine with a total length of 1.8 m. It was fabricated  by multi-layer lamination. The channel consists of three different layers, each with the desired pattern. The first and third layer were made of a 140-µm thick Cyclic Olefin Copolymer (COC) substrate (mcs foil 011, Topas, microfluidic ChipShop, Germany), while the central layer was made from a 127-µm thick Pressure Sensitive Adhesive (PSA) substrate (ARcare® 8939 white PSA, Adhesive Research, Ireland). These microfluidic channel layers were cut using a Graphtec cutting Plotter CE6000-40 (CPS Cutter Printer Systems, Spain). 

The PDMS micropumps were degassed in an RVR003H-01 vacuum chamber (Dekker Vacuum Technologies, USA) at a pressure of 200 Pa below atmospheric pressure for 5, 15 and 30 minutes for each different pump type. Then, the pumps were vacuum-packed individually in an SV-204 vacuum sealer (Sammic, Spain) to store them in an airless environment until use. 

During the experiment, the pump was removed from the vacuum sealer and stucked in the outlet of the a microchannel using a rectangular shaped PSA (146-$\mu$m thick ARcare® 90880, Adhesive Research, Ireland). After waiting for three minutes, a red food coloring solution in water {($\eta_s=1\text{ mPa s}$)} was added to the device’s inlet. The advance of the fluid front was recorded on video (OnePlus 6T rear cameras: 16 + 20 megapixels) at 30 frames per second.

In order to determine the position of the fluid front as a function of time, the video data was examined by timing the fluid front's passage over each 30 mm mark on the serpentine channel.

\subsection{Experiments with air-impermeable coated porous media pumps}
To demonstrate that the mathematical model agrees with the experimental results, we have performed three different experiments of different degassing times using a porous media pump (made by PDMS) with air-impermeable coating (composed by epoxy resin). In all cases we observed an asymptotic behavior of the front movement as a function of time, stopping the front in a given position $x_0$ as predicted by the model.

\begin{figure} [h]
    \centering
    \includegraphics[width=0.5\textwidth]{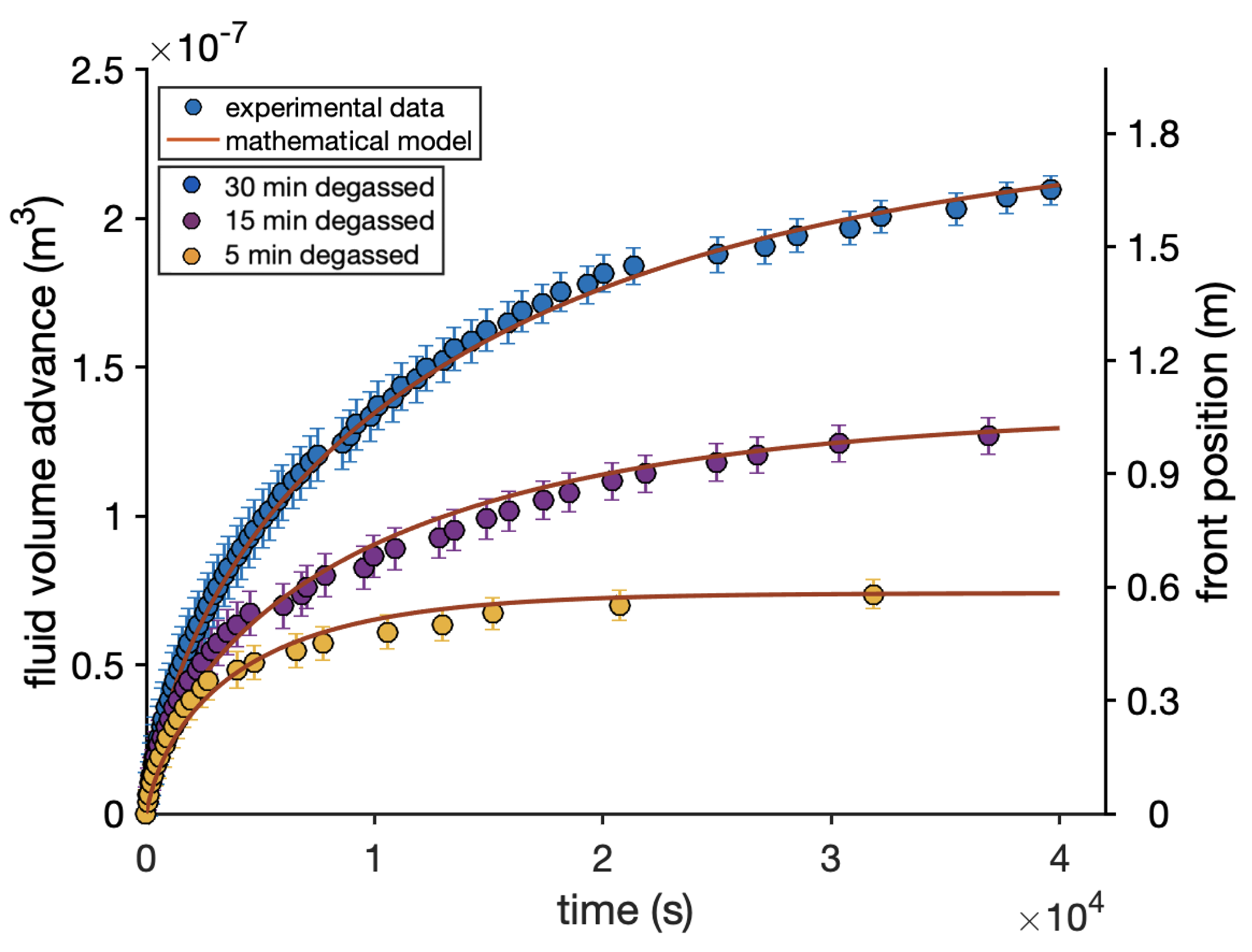}
    \caption{Volume (left) or front position (right) of fluid advance as a function of time moved by the pressure difference generated by a porous pump with an impermeable coating. The pump were degassed 5, 15 and 30 minutes, obtaining three different datasets. In this particular hull, the pump is composed by a PDMS cartige with an epoxy resin coating. The dots correspond to experimental data points and the solid line to the model proposed in Equation (\ref{eq.xcot}).}
    \label{fig.volepox}
\end{figure}

We obtained a dataset for the coated pump subjected to three different degassing times (5, 15, and 30 minutes), along with their corresponding errors.

This dataset has been plotted in Figure \ref{fig.volepox}, where it is also shown the curves given by Equation (\ref{eq.xcot}) for each of the three different degassing times.

\begin{table}[h]
\centering
\begin{tabular}{l|ccc}
$t_d$ (min)              & 30                    & 15                      & 5             \\              \hline
$x_0$ (m)                & 1.81$\pm$0.01                  & 1.05$\pm$0.01                    &  0.59$\pm$0.01   \\
$V_0=V_{des}$ (10$^{-7}$ m$^3$)  & 2.29$\pm$0.01                  & 1.33$\pm$0.01                    &  0.75$\pm$0.01  \\
$R_1$ (10$^{11}$ $\text{Pa}\:\text{s}\:\text{m}^{-3}$)     & 10$\pm$3       & 6$\pm${4}        &  7$\pm${3}       \\
$\tau$ (10$^{3}$ s)      & 21.0$\pm${0.9}                   & 14.9$\pm${1.8}                      & 6.1$\pm$0.8             \\             
$P_0$ (Pa)               & 69$\pm$7                    & 33$\pm$9                      & 29$\pm$9                \\          
$t_{co}$ (10$^{3}$ s)    & 1.3$\pm$0.8                   & 1.1$\pm${1.1}                      & 1.8$\pm${1.8}                              
\end{tabular}
\caption{Table showing the characteristic parameters in the dynamic description of air-impermeable coated porous media pumps for a {degassed time, $t_d$,} 5, 15 and 30 minutes.}
\label{tab.coated}
\end{table}

Table \ref{tab.coated} shows the characteristic variables of each coated pump. $V_0$ (or $x_0=V_0/S_s$) are parameters that have been obtained experimentally by observing at which point the front stops, while the parameters $R_1$ and $\tau$ have been obtained from the least squares fit of the data with Equation (\ref{eq.xcot}). {$P_0$, has been obtained using the expression for the final front position shown in Equation (\ref{eq.finalV}). We have to take into account that, as the effective surface of the pump does not change, $R_1$ must be constant for all the three different experiments}. {Parameter $\xi_s$, which is defined in Equation \eqref{eq.rsamp}, is a known parameter since it only depends on the sample viscosity and the channel geometry.} In table \ref{tab.coated} it can bee seen an increase of long-term front position ($x_0$), time constant ($\tau$) and initial pressure ($P_0$) when increasing the degassing time.

It can be seen that the crossover time ($t_{co}$) is much smaller than the characteristic time constant ($\tau$). In this case, the quadratic term of Equation (\ref{eq.xcot}) dominates the behavior of the system. {Then, the volume advance of the liquid sample through the microchannel can be approximated as:
\begin{eqnarray}
    V_{coated}(t) \approx V_0\left(1-e^{-\frac{t}{\tau}}\right)^{\frac{1}{2}}.
\end{eqnarray}}

\begin{figure} [h]
    \centering
    \includegraphics[width=0.4\textwidth]{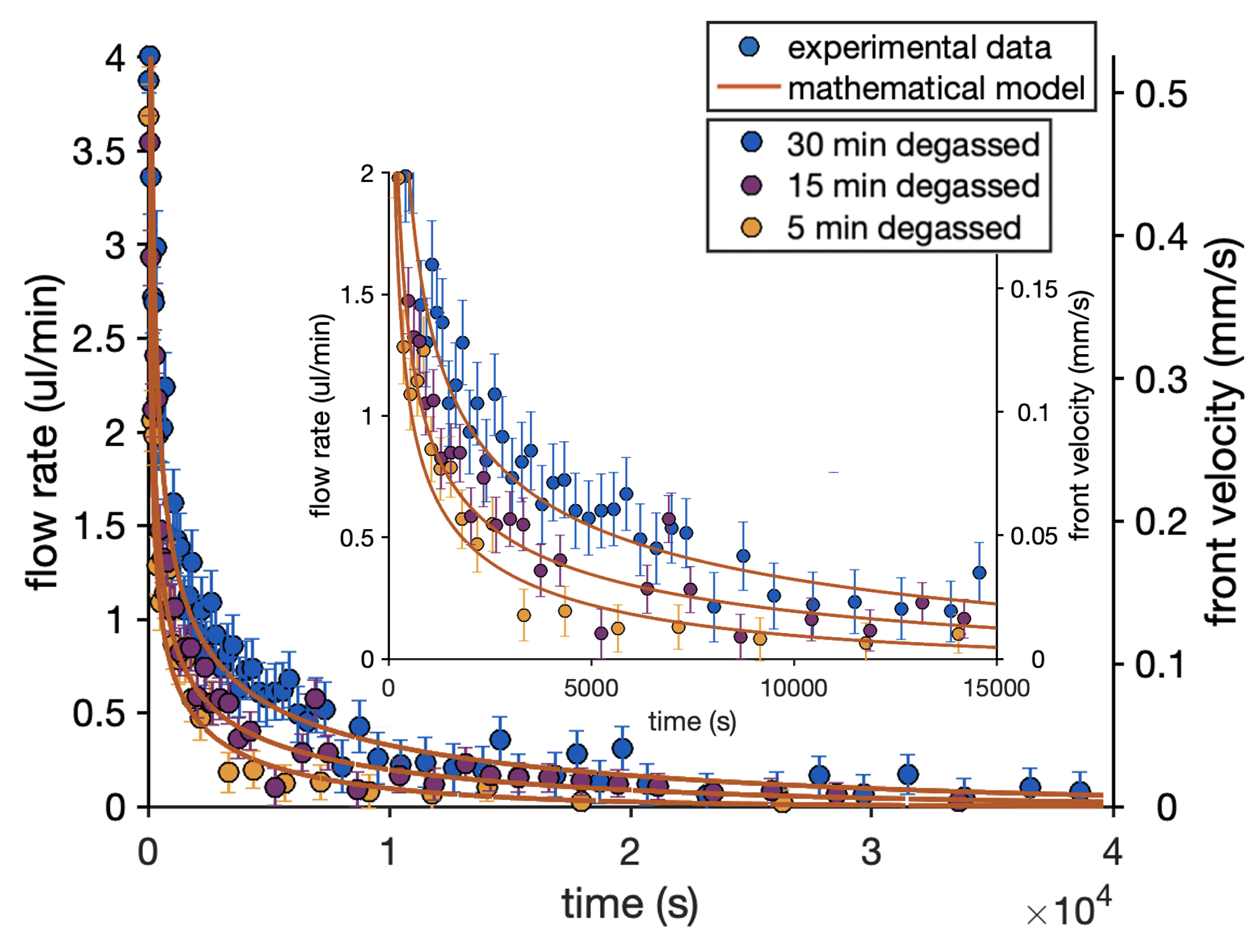}
    \caption{Flow rate (left) or front velocity (right) as a function of time moved by the porous media pump with an impermeable coating. The pump were degassed 5, 15 and 30 minutes, obtaining three different datasets. The dots correspond to experimental acquisitions and the solid line to the model proposed in Equation (\ref{eq.flowcoated}). The figure shows an enlargement of the most relevant area.}
    \label{fig.coatedflow}
\end{figure}

Figure \ref{fig.coatedflow} shows the numerical derivative of the experimental volume (front position) dataset, the flow rate (front velocity) as a function of time that the pump is able to generate. The mathematical model predicting the flow rate of this type of pump, as represented in Equation (\ref{eq.flowcoated}), is also included in this figure.

It can be observed that the longer the pumps are degassed, the greater the air volume they can absorb and the higher the pressure difference they can generate. This is because, as the degassing time increases, the amount of air removed from the PDMS is greater, so the pump tends to recover more air than at lower degassing times. This same effect can also be observed in the flow rate exerted by each pump, obtaining much higher initial flow rates for pumps that have been degassed for a longer time.


We can observe, in all cases, that the mathematical model concerning porous media pumps with impermeable coating aligns well with the experimental results.

\subsection{Experiments with uncoated porous media pumps}

In order to validate the general mathematical model, considering all the resistances, experiments using porous media pumps without any coating (without epoxy resin) were performed. In order to be able to compare this experiment results with the previous case, the pump was degassed at the same pressure and time conditions as in the coated case, obtaining the corresponding dataset with their experimental errors. These experiments exhibited similar behavior to the coated case, though the efficiency decreased significantly due to the absorption of air by the outer surface.

\begin{figure} [h]
    \centering
    \includegraphics[width=0.5\textwidth]{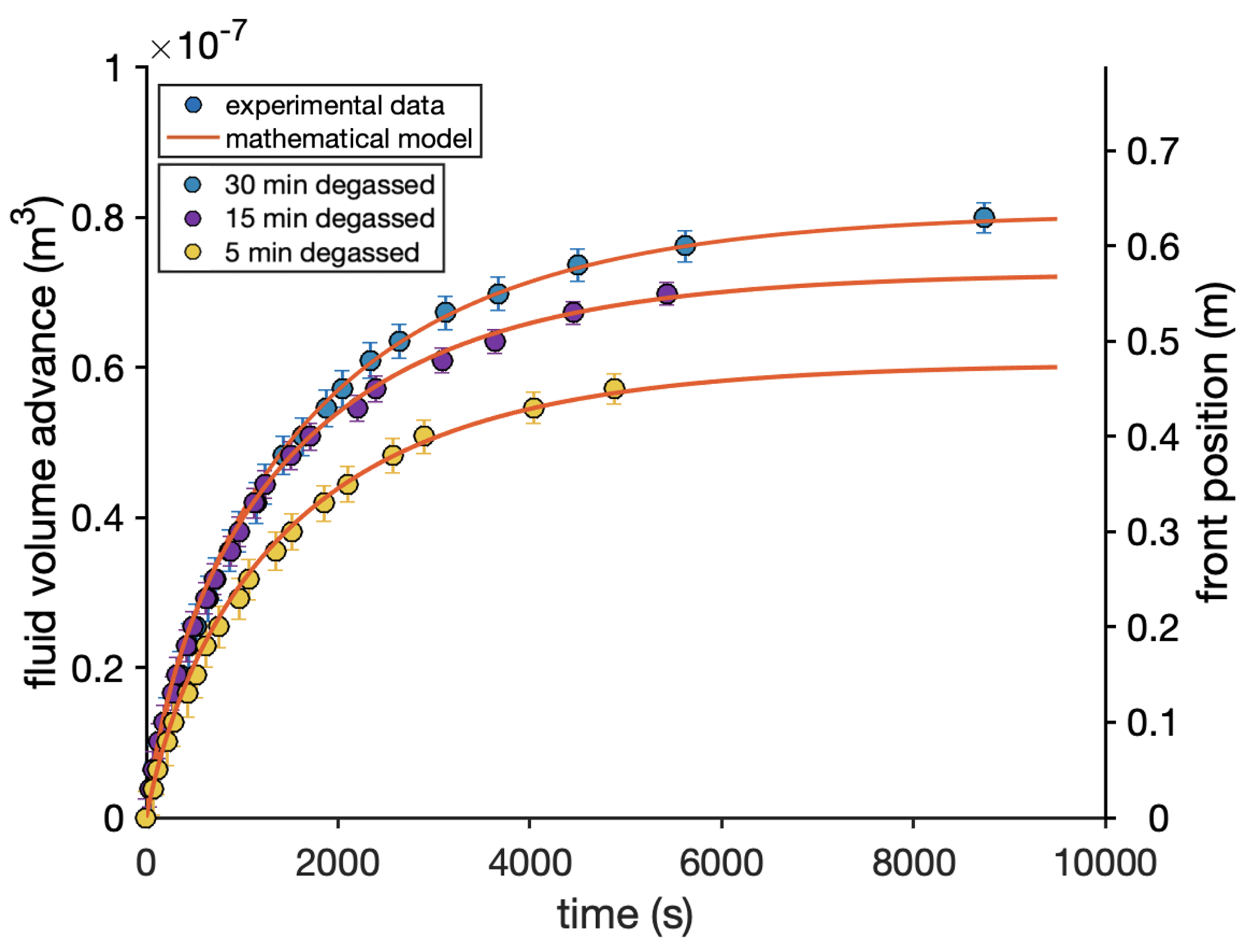}
    \caption{Volume (left) or front position (right) of fluid advance as a function of time moved by the pressure difference generated by a porous media pump without any coating. The pump were degassed 5, 15 and 30 minutes, obtaining three different datasets. In this particular hull, the pump is composed by a PDMS cartige. The dots correspond to experimental acquisitions and the solid line to the model proposed in Equation (\ref{eq.xcot}).}
    \label{fig.volnocoat}
\end{figure}

Figure \ref{fig.volnocoat} shows the experimental data obtained by performing the corresponding experiments for uncoated pumps, where the curve given by the mathematical model proposed in Equation (\ref{eq.xcot}) is plotted. These experiments were conducted using exactly the same method as in the previous case.

\begin{table}[h]
\centering
\begin{tabular}{l|ccc}
$t_d$ (min)              & 30                     & 15                     & 5                    \\       \hline
$x_0$ (m)                & 0.63$\pm$0.01                   & 0.58$\pm$0.01                   &  0.49$\pm$0.01 \\
$V_0$ (10$^{-7}$ m$^3$)  & 0.80$\pm$0.01                   & 0.74$\pm$0.01                   &  0.62$\pm$0.01  \\
$R_1$ (10$^{11}$ $\text{Pa}\:\text{s}\:\text{m}^{-3}$)      & 12$\pm$3       & 9$\pm$5       &  11$\pm$2      \\
$\tau$ (10$^{3}$ s)               & 2.3$\pm$0.2                   & 2.1$\pm$0.4                   & 2.1$\pm$0.2       \\
$P_0$ (Pa)               & 107$\pm$22                    & 85$\pm$35                     & 72$\pm$14        \\
$t_{co}$ (10$^{3}$ s)    & 1.6$\pm$1.1                    & 1.1$\pm$1.1                    & 2.5$\pm${1.9}                   \\         
$R_2$ (10$^{11}$ $\text{Pa}\:\text{s}\:\text{m}^{-3}$)      & 6$\pm$2        & 4$\pm$3        &  6$\pm$1   \\
$V_{des}$ (10$^{-7}$ m$^3$)        &  4$\pm$2       & 4$\pm$4       & 4$\pm$1                           
\end{tabular}
\caption{Table showing the characteristic parameters in the dynamic description of PDMS pumps without any coating for a degassed time, $t_d$, 5, 15 and 30 minutes.}
\label{tab.nocoted}
\end{table}

Table \ref{tab.nocoted} shows the characteristic variables of each uncoated pump. These variables were determined using the same procedure as in the coated case, while $R_2$ and $V_{des}$ where found using Eqs. \ref{eq.rpdms} and \ref{eq.vdes} respectively. In this scenario, we also can see an increase of long-term front position $x_0$ and initial pressure $P_0$, but much more slightly than in the coated case. This is because the air absorption by the external faces is very similar in all cases and dominant in the system. This effect is also observed in the time constant, which is very similar in the three experimental datasets.

{Since the crossover time ($t_{co}$) is of the same magnitude as the characteristic time constant ($\tau$), and most of the experimental points lie behind this time threshold, we can approximate the behavior of the volume advance of the liquid sample through the
microchannel as:
\begin{eqnarray}
    V_{uncoated}(t) \approx V_0\left(1-e^{-\frac{t}{\tau}}\right).
\end{eqnarray}}

\begin{figure} [h]
    \centering
    \includegraphics[width=0.4\textwidth]{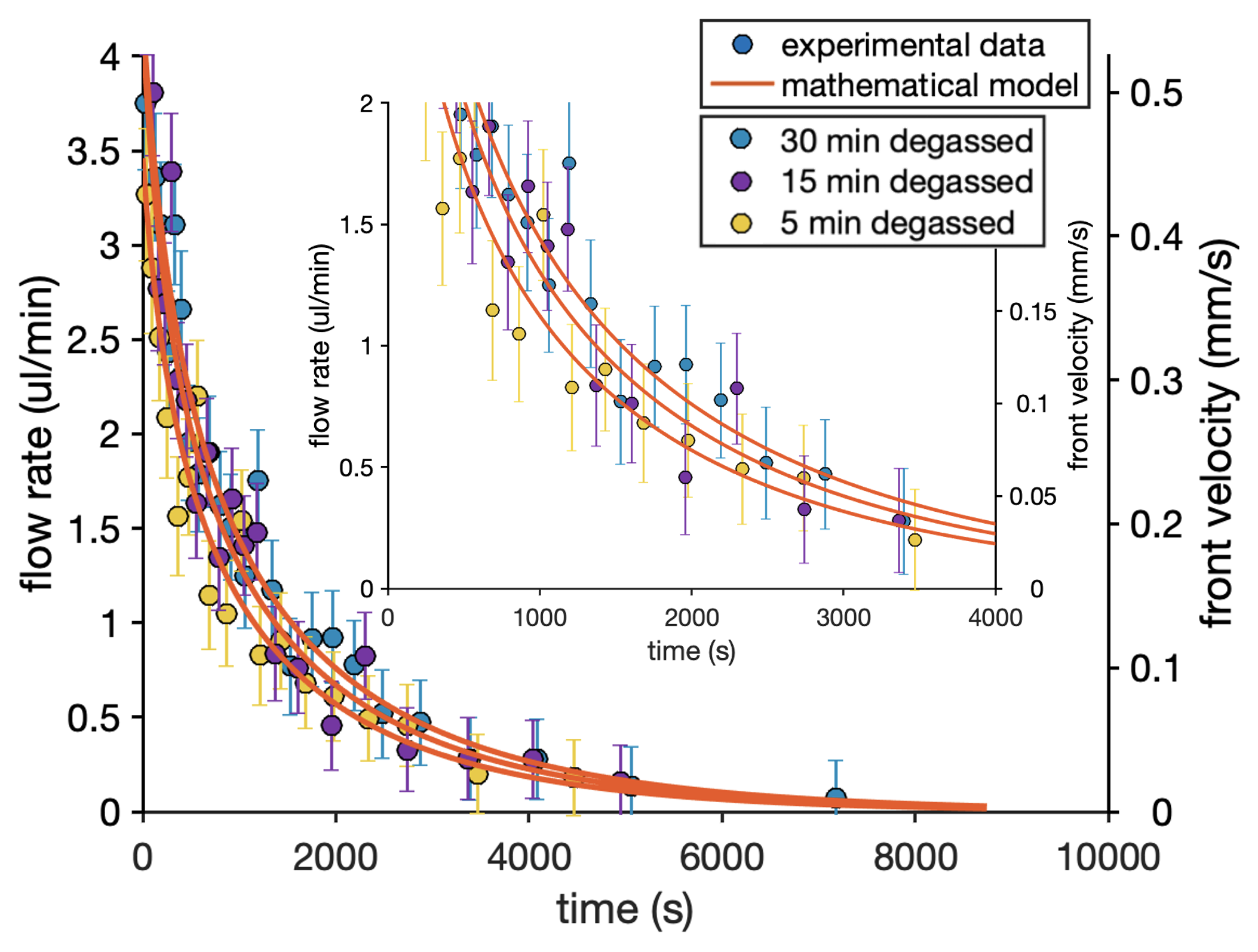}
    \caption{Flow rate (left) or front velocity (right) as a function of time moved by the porous media pump without any air-impermeable coating. The pump were degassed 5, 15 and 30 minutes, obtaining three different datasets. The dots correspond to experimental acquisitions and the solid line to the model proposed in Equation (\ref{eq.flowcoated}). The figure shows an enlargement of the most relevant area.}
    \label{fig.nocoatedflow}
\end{figure}

The flow rate and the front velocity generated by the uncoated pumps can be seen in Figure \ref{fig.nocoatedflow}. In this case, as in the previous one, the flow (velocity) data corresponds to the numerical derivative of the volume data (front position data) as a function of time. In the figure we have also included the theoretical model for the flow rate, represented in Equation (\ref{eq.flowcoated}). See that, for uncoated pumps, the velocity is very similar in all cases, this is also due to the dominant effect of the absorption by the external faces .

The volume of air removed from the pump in the degassing process is much greater than the volume absorbed by the channel. This is due to the fact that the pump has much more facility to absorb air by its external surface than by the surface in contact with the channel due to the resistance of the liquid to be moved.

We can see that, for all experiments with uncoated PDMS pumps, there is a correct agreement between the mathematical model and the experimental results.

\subsection{Comparison between air-impermeable coated and uncoated porous media pumps experiments}

The degassing process for both coated and uncoated pumps is identical. However, due to the absence of coating, more air is removed from the uncoated pumps, which have a larger exposed surface area. The key distinction between the two types of pumps is the air losses that occur through the external surfaces of the uncoated pumps. This effect can be observed by comparing the volume removed during the degassing process ($V_{des}$) in tables \ref{tab.coated} and \ref{tab.nocoted}.

\begin{figure} [h]
    \centering
    \includegraphics[width=0.5\textwidth]{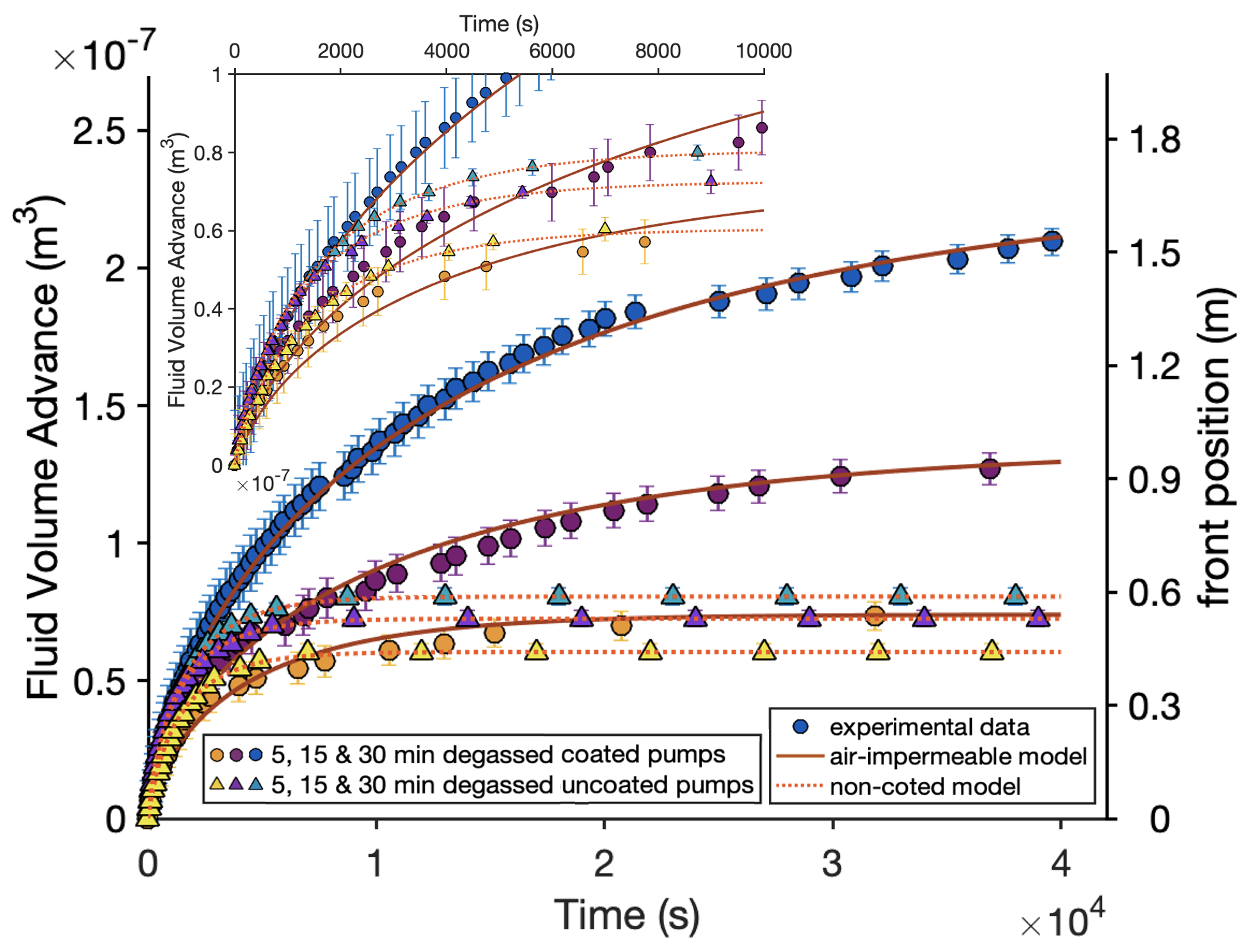}
    \caption{Volume (left) or front position (right) of fluid advance as a function of time moved by both, the air-impermeable porous media pump (circles and solid line) and the uncoated porous media pump (triangles and dotted line) in order to compere their dynamics. The pumps were degassed 5, 15 and 30 minutes in for each case. {The lines correspond to the mathematical model proposed in Equation (\ref{eq.xcot}). The figure shows an enlargement of the short time values. }}
    \label{fig.comparx}
\end{figure}

Figure \ref{fig.comparx} illustrates that, over time, the volume of liquid moved by the coated pump (long-term position) is significantly higher than the one moved by the uncoated ones. {Covering the pump can increase the displaced volume of the fluid sample by almost three times for 30 minutes of degassing. This difference decreases with decreasing degassing time.}

Mathematically, this main difference between the pumps translates into a drastic decrease in the time constant ($\tau$) when the pump is uncoated, compared with the coated case. This phenomenon is described in Equation (\ref{eq.tau}) and its comparison with Equation (\ref{eq.tauimp}). Experimentally, we can see that coated pumps run up to ten times longer than their uncoated counterparts, which corroborates this theory.

Another important effect of pressure losses on the uncoated pumps is evident in the degassing times. While the coated pumps show a larger difference in performance between degassing intervals, uncoated pumps exhibit a reversed pattern. This is because, as more air is removed from the uncoated pumps, they reabsorb air more quickly through their external surfaces. For instance, in coated pumps, the difference in long-term position between 5 and 15 minutes of degassing is smaller than that between 15 and 30 minutes. However, when external air losses occur, this pattern reverses, as can be seen in Figure \ref{fig.comparx}.

\begin{figure} [h]
    \centering
    \includegraphics[width=0.5\textwidth]{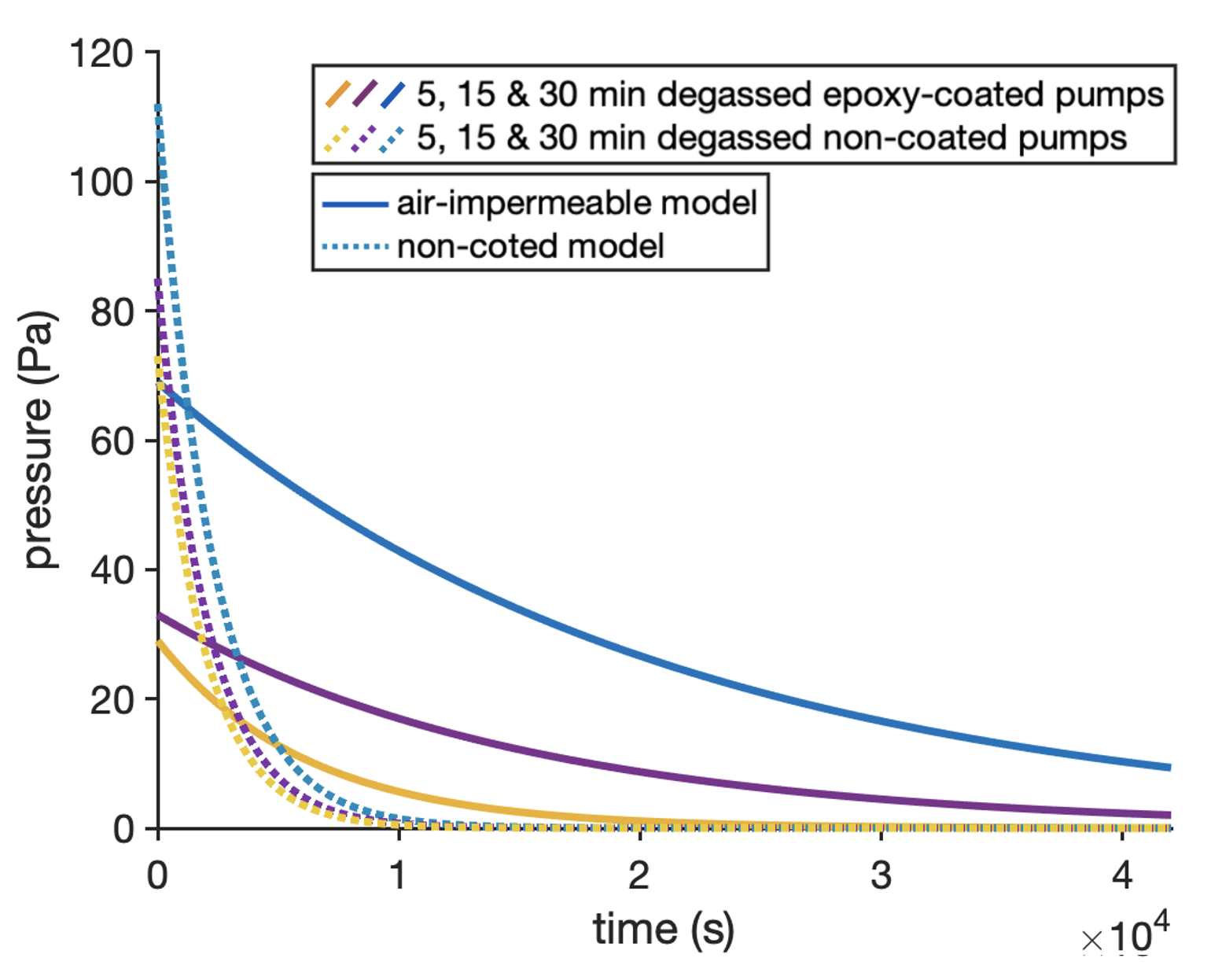}
    \caption{Time-dependent pressure difference responsible for moving the fluid through the channel for both, the air-impermeable porous media pump (solid line) and the uncoated porous media pump (dotted line), using Equation (\ref{eq.difP}) in order to compare their dynamics. The pump were degassed 5, 15 and 30 minutes in for each case. }
    \label{fig.comparep}
\end{figure}

The pressure exerted by each pump as a function of time can be seen in Figure 9. The effect of absorption through the external faces is clearly evident, as the initial pressure of the uncoated pumps is much higher, but it decreases rapidly. In contrast, for the coated pumps, the pressure decrease is much more gradual.

\section{CONCLUSIONS}

In this article, a mathematical model based on fluid dynamics, that correctly describes the behavior of a liquid inside a channel moved by self-powered porous media pump, has been presented. This model allows to use porous media pumps in the field of microfluidics, knowing with precision the behavior of the liquid that it moves.

The mathematical model consists of a system of fluidic resistances, two in series and one in parallel, capable of describing the behavior of any porous medium that is previously degassed. Equation (\ref{eq.xcot}) shows the volume of liquid that the pump is capable of moving as a function of time, the geometry of the microchannel, the viscosity of the fluid and the properties of the porous medium.

When the external surface of a porous media pump is sailed with an air-impermeable material coating, it is not able to absorb air through the environment avoiding pressure losses. As a result, all the air removed during the degassing process is reabsorbed through the non-coated surface of the pump in contact with the channel. The absorption of channel's air displaces the liquid sample. On the other hand, in the case of the uncoated pump, the air is reabsorbed by the entire surface of the pump. Although the surface in contact with the channel also absorbs air, the majority is reabsorbed through the external surface due to the resistance posed by the liquid within the channel.

The difference between coated and uncoated pumps results, mathematically, in an increase in the characteristic time constant ($\tau$) when coated. Obtaining an expression for this constant in Equation (\ref{eq.tau}) and its comparison with Equation (\ref{eq.tauimp}).

Furthermore, we have experimentally tested the behavior of this type of pumps using PDMS as the porous medium and epoxy resin as the air-impermeable coating. We have observed and discussed the differences between the two types of pumps. The most relevant difference is that, with the same degassing conditions, the coated pumps have significantly increased the volume of liquid absorbed in the channel with respect to the uncoated ones. We have also successfully verified that the mathematical model proposed in this paper agrees with the data obtained experimentally.

In conclusion, this paper demonstrates the effectiveness of mathematical models in characterizing porous media pumps for flow control in long-term autonomous microsystems.

\begin{acknowledgments}
We acknowledge financial support from ‘‘Red Temática MIFLUNET’’.
AB-C and AH-M acknowledge support from ‘‘Ministerio de Ciencia e Innovación’’ (Spain) under project PID2022-137994NB-100 and ‘‘AGAUR’’ (Generalitat de Catalunya) under project 2021 SGR 00450.
YA-B, FB-L and LB-D acknowledge funding support from “Ministerio de Ciencia y Educación de España” under grant PID2020-120313GB-I00/AIE/10.13039/501100011033 and from Basque Government under ‘‘Grupos Consolidados’’ with Grant No. IT1633-22.

\end{acknowledgments}

\nocite{*}
\bibliography{aipsamp}

\end{document}